\documentclass[aps,prl,twocolumn,superscriptaddress]{revtex4-2}
\usepackage{empheq}
\usepackage{graphicx}
\usepackage{epsfig}
\usepackage{amsmath,amsthm}
\usepackage{color}
\usepackage{verbatim}
\usepackage{ulem}
\usepackage{siunitx}
\usepackage{lineno}

\begin{document}

\preprint{unpublished manuscript}
%\linenumbers

%\bibliographystyle{prbsty}

\title{Current-driven writing process in antiferromagnetic Mn$_2$Au for memory applications}

\author{S.\,Reimers $^\dagger$}
\affiliation{Institut f\"{u}r Physik, Johannes Gutenberg-Universit\"{a}t Mainz, 55099 Mainz, Germany}
\author{Y.\,Lytvynenko $^\dagger$}
\affiliation{Institut f\"{u}r Physik, Johannes Gutenberg-Universit\"{a}t Mainz, 55099 Mainz, Germany}
\affiliation{Institute of Magnetism of the NAS of Ukraine and MES of Ukraine, 03142 Kyiv, Ukraine}
\author{Y.\,R. Niu}
\affiliation{MAX IV Laboratory, Fotongatan 8, 22484 Lund, Sweden}
\author{E.\,Golias}
\affiliation{MAX IV Laboratory, Fotongatan 8, 22484 Lund, Sweden}
\author{B.\,Sarpi}
\affiliation{Diamond Light Source, Chilton, Didcot, Oxfordshire, OX11 0DE, United Kingdom}
\author{L.\,S.\,I.\,Veiga}
\affiliation{Diamond Light Source, Chilton, Didcot, Oxfordshire, OX11 0DE, United Kingdom}
\author{T.\,Denneulin}
\affiliation{Ernst Ruska-Centre for Microscopy and Spectroscopy with Electrons, Forschungszentrum J{\"u}lich, 52425J{\"u}lich, Germany}
\author{A.\,Kov{\'a}cs}
\affiliation{Ernst Ruska-Centre for Microscopy and Spectroscopy with Electrons, Forschungszentrum J{\"u}lich, 52425J{\"u}lich, Germany}
\author{R.\,E.\,Dunin-Borkowski}
\affiliation{Ernst Ruska-Centre for Microscopy and Spectroscopy with Electrons, Forschungszentrum J{\"u}lich, 52425J{\"u}lich, Germany}
\author{J.\,Bl{\"a}{\ss}er}
\affiliation{Institut f\"{u}r Physik, Johannes Gutenberg-Universit\"{a}t Mainz, 55099 Mainz, Germany}
\author{M.\,Kl{\"a}ui}
\affiliation{Institut f\"{u}r Physik, Johannes Gutenberg-Universit\"{a}t Mainz, 55099 Mainz, Germany}
\author{M.\,Jourdan*}
\affiliation{Institut f\"{u}r Physik, Johannes Gutenberg-Universit\"{a}t Mainz, 55099 Mainz, Germany}
\email{jourdan@uni-mainz.de}

\begin{abstract}
Current pulse driven N{\'e}el vector rotation in metallic antiferromagnets is one of the most promising concepts in antiferromagnetic spintronics. We show microscopically that the N{\'e}el vector of epitaxial thin films of the prototypical compound Mn$_2$Au can be reoriented reversibly in the complete area of cross shaped device structures using single current pulses. The resulting domain pattern with aligned staggered magnetization is long term stable enabling memory applications. We achieve this switching with low heating of $\approx 20$~K, which is promising regarding fast and efficient devices without the need for thermal activation. Current polarity dependent reversible domain wall motion demonstrates a N{\'e}el spin-orbit torque acting on the domain walls.     
\end{abstract}

%\maketitle

%\keywords{Spin polarization, momentum microscopy,  photoemission electron spectroscopy, ARPES}
%\pacs{79.60-i, 73.20-r, 75.25-j, 75.70-i}

\maketitle

* email: Jourdan@uni-mainz.de\\
$^\dagger$ These authors contributed equally to this work.

\section{INTRODUCTION}
Novel storage concepts in spintronics based on antiferromagnets (AFMs) propose to encode information by the direction of the alignment of the staggered magnetization or N{\'e}el vector \cite{Mac11,Jun18,Bal18,Jung18}. This approach takes advantage of the intrinsically fast THz dynamics of AFMs \cite{Kam11} and their stability against external magnetic fields (see e.\,g.\,\cite{Sap18}). The interest in such concepts has grown strongly with the prediction of a so called N{\'e}el spin-orbit torque (NSOT) in metallic collinear AFMs with a specific symmetry, which is expected to rotate the spins of the AFM sublattices in a perpendicular orientation with respect to a driving bulk current \cite{Zel14}. 
 
Correspondingly, within the first generation of experiments on both known compounds with the required structure and AFM ordering, CuMnAs and Mn$_2$Au, current pulse induced reversible resistance modifications were reported as evidence for NSOT driven N{\'e}el vector rotation \cite{Wad16, Grz17, Mat19, Bod18, Mei18, Zho18}. However, it later became clear that alternative mechanisms such as current pulse induced heat effects \cite{Chi19, Zin19}, electromigration \cite{Mat20}, and rapid quenching induced structural and magnetic modifications \cite{Kas20} can result in resistance modifications similar to the ones reported as well. Additionally, external strain supported N{\'e}el vector manipulation in Mn$_2$Au was demonstrated \cite{Che19,Gri22}. Thus, microscopic investigations showing the intended current driven alignment of the staggered magnetization directly are essential. Up to now, for CuMnAs as well as for Mn$_2$Au, only minor current induced modifications were observed, i.\,e.\,only a small fraction of the antiferromagnetic domains switched with limited reversibility and stability \cite{Grz17,Bod19}. This is neither sufficient regarding the identification of the switching mechanism as local inhomogeneities are involved nor does it demonstrate the applicability of current induced N{\'e}el vector switching for memory devices.

Furthermore, microscopic imaging provides direct insights into the current driven mechanisms of N{\'e}el vector reorientation \cite{Wad18}. This recently attracted renewed interest, as in NiO/Pt bilayers thermomagnetoelastic coupling effects were observed, which act in a very similar way as spin-orbit torques on the N{\'e}el vector \cite{Mee21}. Thus, the experimental demonstration of a current induced NSOT is important.

Here we show current pulse induced complete, remanent, and reversible N{\'e}el vector switching of Mn$_2$Au(001) thin films patterned in cross-structures, which is compatible with spintronics applications such as magnetic random access memory (MRAM). A current polarity dependence demonstrates an NSOT acting on AFM domain walls. Studying different pulse lengths, we show that in contrast to related experiments on Mn$_2$Au \cite{Mei18}, for our epitaxial thin films thermal activation is not necessary for switching, which is important for fast and energy efficient memory applications regarding potential ultrafast applications.

\section{Results}

Mn$_2$Au is a metallic antiferromagnet with a high N{\'e}el temperature above \SI{1000}{\K} \cite{Bar13}. The compound has a tetragonal crystal structure with two equivalent $\langle 110 \rangle$ easy axes in the (001)-plane and a strong out-of-plane magnetic anisotropy. 
 
We investigate epitaxial Mn$_2$Au(001) thin films with a thickness of \SI{45}{\nm} grown on an epitaxial double buffer layer of \SI{13}{\nm} of Ta(001) on \SI{20}{\nm} of Mo(001) on MgO(001) substrates (see Supplementary Information). The samples are capped with \SI{2}{\nm} of SiN$_x$ and patterned by optical lithography and Argon ion beam etching.   

The largest NSOT is expected for the current directions parallel and antiparallel to the N{\'e}el vector \cite{Zel14,Sal19}. Correspondingly, we patterned a Mn$_2$Au(001) thin film in a cross structure oriented parallel to the easy $\langle 110 \rangle$ directions, which allows to send current pulses both parallel and perpendicular to the axis along which the N{\'e}el vector is aligned. We obtain a repeatable complete switching of the N{\'e}el vector orientation in the central area $(\SI{10}{\micro m} \times \SI{10}{\micro m}$) of the cross by applying current pulses alternating between the two orthogonal directions. Fig.\,1 shows X-ray magnetic linear dichroism - photoelectron emission microscopy (XMLD-PEEM) images obtained after subsequent pulses, in which the \SI{90}{\degree} reorientation of the N{\'e}el vector shows up as alternating dark and white contrast of the central area.

\begin{figure*}
\includegraphics[width=2.0\columnwidth]{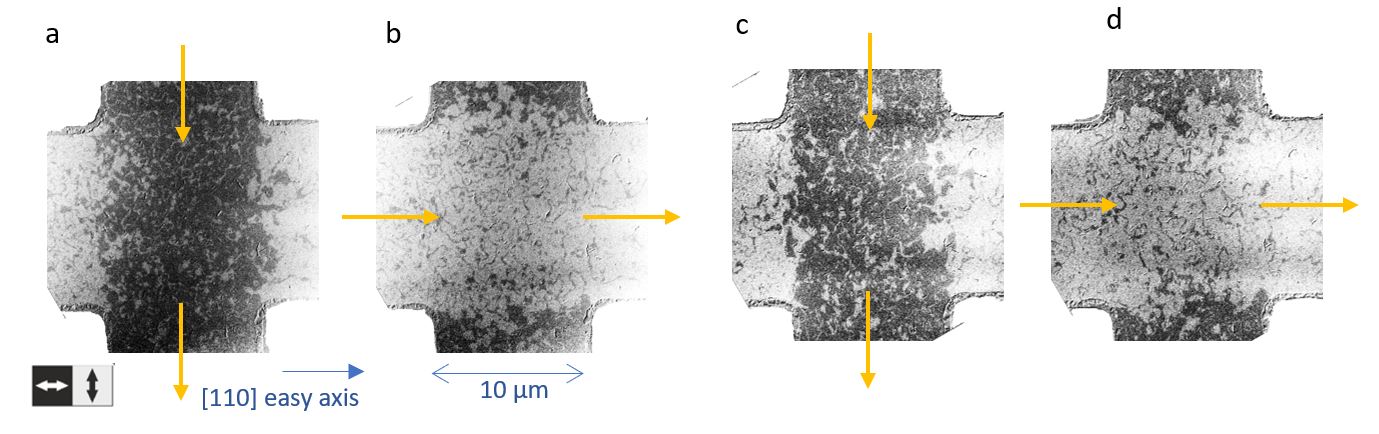}
\caption{\label{Fig1} 
{\bf XMLD-PEEM images of reversible N{\'e}el vector reorientation, with current parallel easy axes.} The images show the orientation of the N{\'e}el vector of Mn$_2$Au(001) thin films after sending current pulses with different length and direction (yellow arrows) through a patterned cross structure oriented parallel to the easy $\langle 110 \rangle$ directions. The dark regions correspond to a horizontal, the bright regions to a vertical alignment of the N{\'e}el vector as indicated below panel {\bf a} by the double arrows. {\bf a, b}: After 100 pulses of 1~ms length each with a current density of $J=2.6 \times 10^{11}$~A$/$m$^2$.  {\bf c,d}: After 1 bipolar pulse of 10~$\mu$s length with a current density of $J=3.0 \times 10^{11}$~A$/$m$^2$.}
\end{figure*}

For the investigation of the role of current heating induced thermal activation \cite{Mei18}, we compare current pulse trains with different numbers and lengths, ranging from a train of 100 pulses with a lengths of \SI{1}{\ms} down to a single \SI{10}{\micro s} pulse. The panels {\bf a} and {\bf b} of Fig.\,1 show an example of complete switching after a train of 100 current pulses with a length of \SI{1}{\ms} each (off time after each pulse \SI{10}{\ms}). The panels {\bf c} and {\bf d} show very similar switching obtained after a single bipolar current pulse with a length of \SI{10}{\micro s} only. Although only a single pulse instead of 100 pulses was used and the pulse length was 100 times shorter, the required current densities to obtain complete N{\'e}el vector rotation are very similar with $J_{{\rm 1ms}}\approx 2.6 \times 10^{11}$~A$/$m$^2$ and $J_{10{\rm \mu s}}\approx 3.0 \times 10^{11}$~A$/$m$^2$ (and $J_{100{\rm \mu s}}\approx 2.7 \times 10^{11}$~A$/$m$^2$, not shown). These current densities and pulse lengths result in significantly different temperature increases of the Mn$_2$Au thin film due to ohmic heating with $\Delta T(1~{\rm ms}) \approx 70$~K, $\Delta T(100~{\rm \mu}$s)$\approx 45$~K, and $\Delta T(10~{\rm \mu}$s)$\approx 20$~K, as determined from ex-situ reference resistance measurements during application of current pulses (see Supplementary Information).

Next we investigate how the switched area of the sample depends on the current density. We patterned a single stripe along an easy $[$110$]$ direction of a Mn$_2$Au thin film and applied \SI{1}{\ms} current pulses with increasing current density, as shown in Fig.\,2. The single stripe geometry ensures a homogeneous distribution of the current density. The transition from the beginning of N{\'e}el vector reorientation to complete switching with increasing pulse current density occurs within a range of $\approx \SI{20}{\percent}$ of the maximum required current. 

\begin{figure*}
\includegraphics[width=2.0\columnwidth]{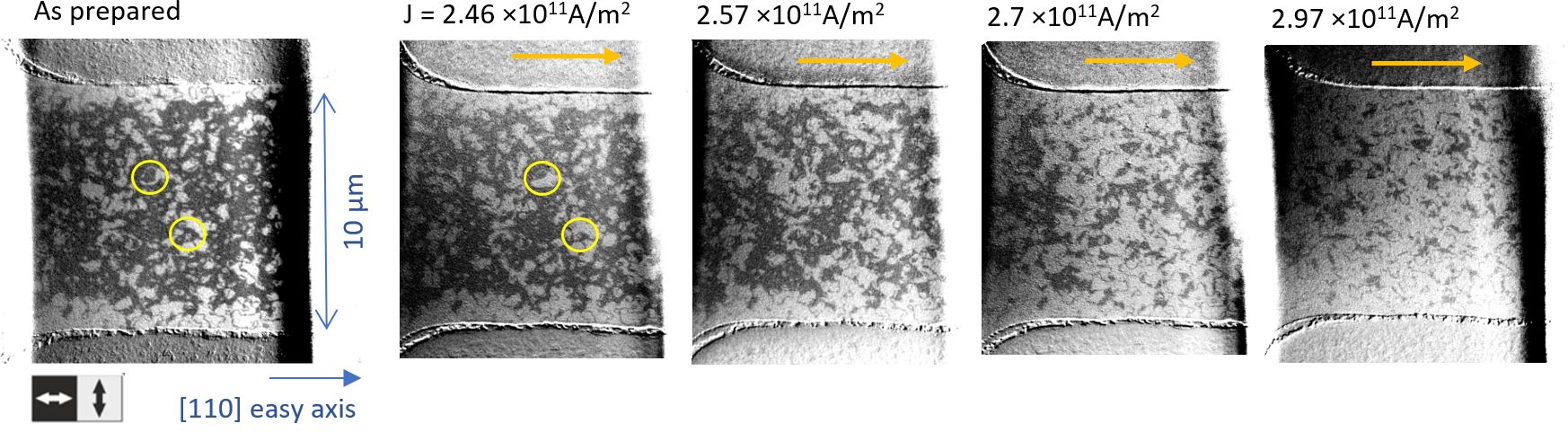}
\caption{\label{Fig2} 
{\bf XMLD-PEEM images of N{\'e}el vector reorientation, with current along easy axis.} The images show the orientation of the N{\'e}el vector of Mn$_2$Au(001) thin films as grown and after sending current pulses (length 1~ms) with increasing amplitude along an easy $\langle 110 \rangle$ direction through a patterned stripe structure. The yellow arrows indicate the current directions.} 
\end{figure*}

The left panel of Fig.\,2 shows the initial domain configuration, in the as grown state after patterning. It is characterised by a preferential alignment of the N{\'e}el vector parallel to the stripe, which originates from patterning induced anisotropy \cite{Rei22}. With increasing amplitude of the current pulses (panels from left to right in Fig.\,2), the first reorientation of the N{\'e}el vector appears in some central regions of the stripe indicated by yellow circles with a current density of $J_{{\rm 1ms}}= 2.46 \times 10^{11}$~A$/$m$^2$. With further increasing current, the switched area increases homogeneously distributed over the patterned area, consistent with the absence of patterning induced current inhomogeneities. In the same experiment with an MBE grown Mn$_2$Au(001) thin film \cite{Bom20} and shorter current pulses (\SI{20}{\micro s}) very similar results were obtained, showing the robustness of the switching properties.

In order to identify if the switching is driven by NSOT, we investigated the current direction and polarity dependence of the N{\'e}el vector reorientation. This allows us to compare the experimental results of changes to the AFM domains and domain wall positions to theoretical predictions for NSOT \cite{Zel14,Sal19}. Current pulses were applied along a hard $[$100$]$ direction of a Mn$_2$Au(001) thin film using a single patterned stripe aligned along this direction. In this geometry, no net alignment of the N{\'e}el vector was obtained in the central region of the patterned stripe. However, we observed a partial current polarity dependent reversible reorientation. Some of these regions which switch for positive current polarity from vertical to horizontal alignment are indicated by the red circles in Fig.\,3. Examples of other regions which switch in the opposite direction are indicated by yellow circles. The green circles in panel {\bf f} show regions, in which the current was flowing preferentially along an easy $\langle 110 \rangle$-direction resulting in a local alignment of the {N\'e}el vector corresponding to the behavior discussed above (Figs.\,1 and 2).

\begin{figure*}
\includegraphics[width=2.0\columnwidth]{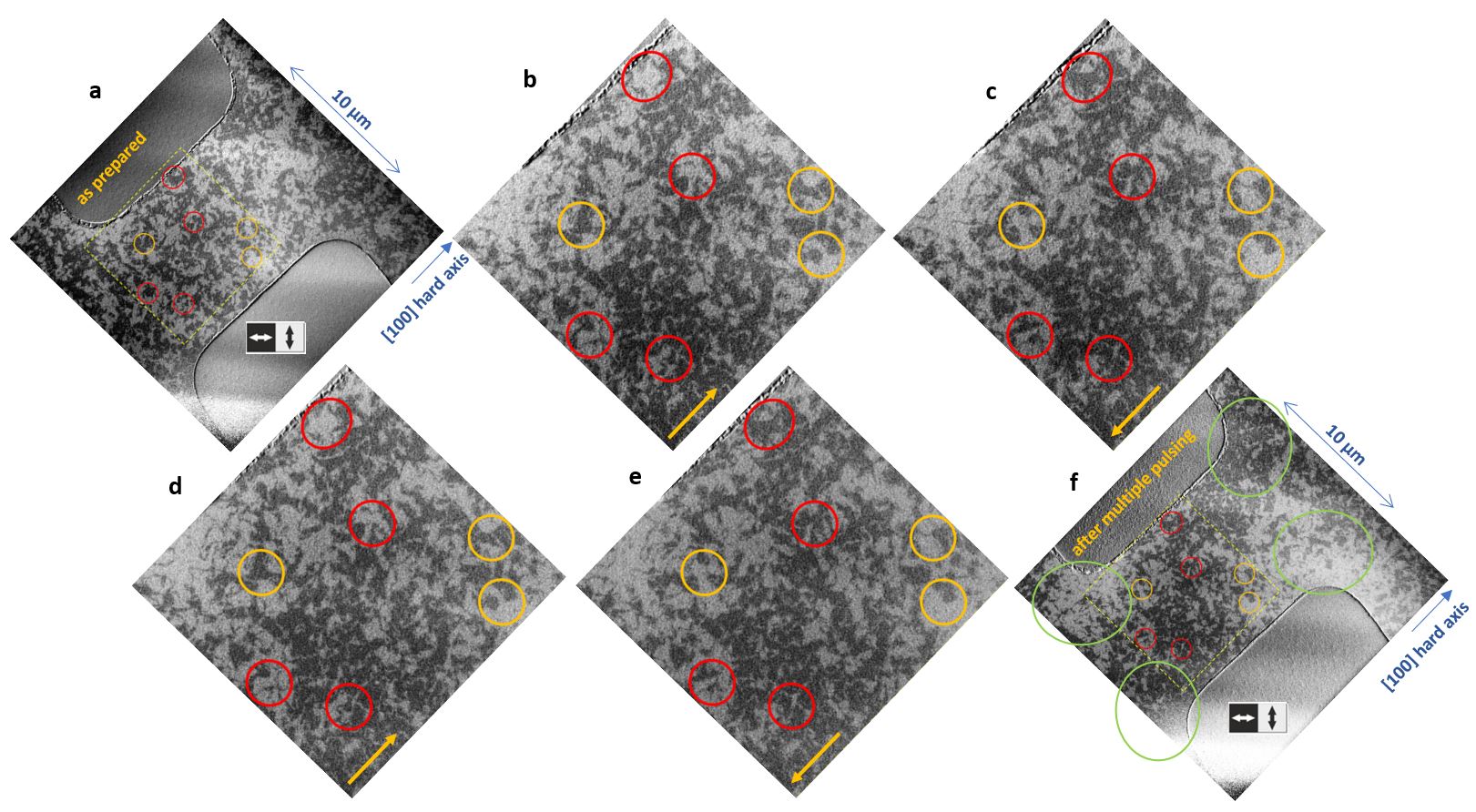}
\caption{\label{Fig3} 
{\bf XMLD-PEEM images of N{\'e}el vector reorientation, with current parallel hard axis.} The images show the orientation of the N{\'e}el vector of Mn$_2$Au(001) thin films as grown (panel {\bf a}) and after sending current pulses (length \SI{1}{\ms}, amplitude $J=3.6 \times 10^{11}$~A$/$m$^2$) with alternating polarity indicated by yellow arrows along the hard $[$100$]$ direction through a patterned stripe structure with panels {\bf b} to {\bf e} showing the region indicated by the yellow dashed square in panel {\bf a} with increased magnification. Panel {\bf f} shows the full field of view after the last current pulse.} 
\end{figure*}  

Finally, we determine the change of the sample resistance associated with the N{\'e}el vector reorientation discussed above. For this, an 8-terminal shaped device as shown in panel {\bf b} of Fig.\,4 was patterned. We applied bipolar current pulses alternating between the two orthogonal easy $\langle 110 \rangle$ directions, as done previously for the 4-terminal cross structures shown in Fig.\,1. With the 8-terminal geometry, both the longitudinal $R_{\rm long}$ as well as the transversal $R_{trans}$ sample resistance can be probed after each current pulse as shown in Fig.\,4. Pulsing with the same current densities as required for the observation of the N{\'e}el vector reorientation by X-PEEM, we observed alternating longitudinal (panel {\bf a}) as well as transversal (see Supplementary Material) resistance values. Both magnetoresistance signals are as expected of similar magnitude, slightly inhomogeneous N{\'e}el vector alignment in the central area of the 8-terminal cross might explain the remaining differences.
\begin{figure}
\includegraphics[width=1.0\columnwidth]{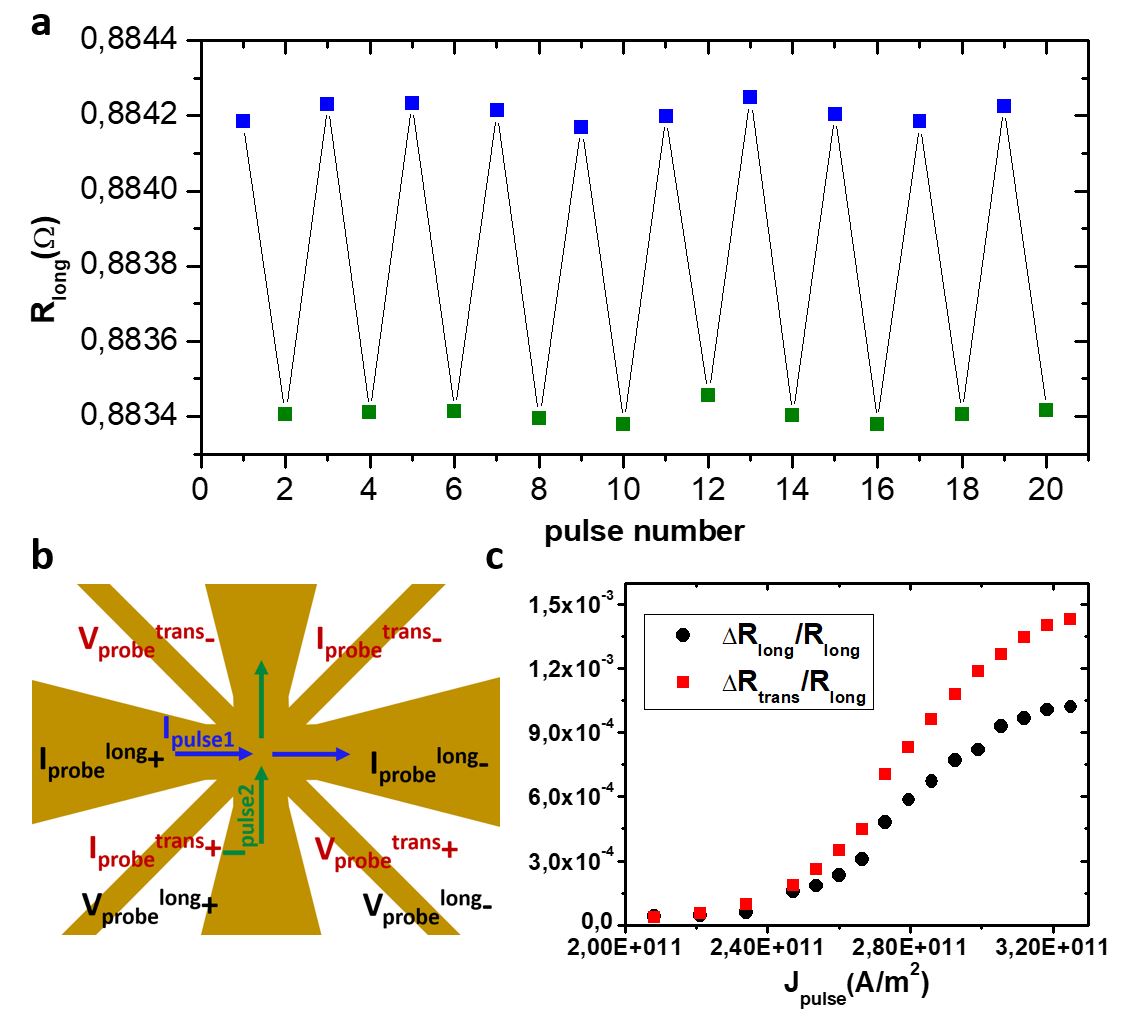}
\caption{\label{Fig4} 
{\bf Switching induced anisotropic magnetoresistance.} Panel {\bf a} shows the alternating longitudinal resistance of a Mn$_2$Au thin film patterned in the geometry shown in panel {\bf b}. The longitudinal resistance was measured after each 1~ms current pulse with $J_{\rm pulse}=3 \times 10^{11}$~A$/$m$^2$ applied alternately in perpendicular directions as indicated by the blue and green arrows in panel {\bf b}. The color of the data points corresponds to the arrows indicating the current direction. Panel {\bf c} shows the dependence of the longitudinal $\Delta R_{\rm long}$ as well as of the transverse $\Delta R_{\rm trans}$ resistance changes on the pulse current density $J_{\rm pulse}$.}
\end{figure}  
We obtained a maximum $\Delta R_{\rm long}/R_{\rm long} \simeq 1\times 10^{-3}$, which is 
consistent with the low temperature anisotropic magnetoresistance value of AMR$_{\rm Mn_2Au}\simeq -1.5\times10^{-3}$, which we obtained previously by aligning the N{\'e}el vector with a 50~T magnetic field pulse \cite{Bod20}.
Furthermore, the negative AMR of Mn$_2$Au (i.\,e.\,$\rho_{\perp} > \rho_{\parallel}$) is 
consistent with the larger longitudinal resistance in Fig.\,4 (panel {\bf a}) associated with pulse current direction $I_{\rm pulse1}$ (blue data points), if the N{\'e}el vector is aligned perpendicular to the current direction as expected for an NSOT acting. 

Furthermore, consistent with the X-PEEM measurements, the electrical signal shows no sign of decay as shown in the Supplementary Information. 

\section{Discussion}

For potential spintronics applications such as antiferromagnetic random access memory (MRAM), long term stability of the N{\'e}el vector aligned states is required. In this framework our Mn$_2$Au(001) thin films are very suitable as the current pulse induced AFM domain configurations perfectly fulfill this requirement. We were able to demonstrate by XMLD-PEEM that even four months after the current pulse N{\'e}el vector alignment shown in Fig.\,2, the aligned AFM domain configuration was preserved (see Supplementary Information). Furthermore, already single current pulses are able to reorient the N{\'e}el within the central area of device cross shaped structures completely and reversibly, i.\,e.\,again fulfilling the requirements for applications. 

For spintronics, the size of the electrical read-out signal associated with N{\'e}el vector reorientation is of major importance. Our work confirms that the maximum obtainable signal from AMR is relatively small so that in antiferromagnetic spintronics other read-out mechanisms, for example via an adjacent strongly exchange-coupled ferromagnetic layer as discussed in reference \cite{Bom21}, need to be considered. Furthermore, our small electrical read-out signals are clearly distinguishable from the current pulse induced large resistance modifications previously obtained investigating various non-NSOT related metals as mentioned in the introduction. 

Regarding the physical mechanism of the current pulse induced N{\'e}el vector reorientation, we compare our experiments with theoretical predictions assuming an NSOT, which is expected to be largest for current direction parallel to the sublattice magnetization and zero for current perpendicular to it \cite{Zel14, Sal19}. Fully consistent with this prediction, we observe a current induced N{\'e}el vector orientation perpendicular to the pulse direction, if the current flows parallel to an easy $\langle 110 \rangle$-axis of Mn$_2$Au (Figs.\,1 and 2).
However, also thermomagnetoelastic coupling effects driven by anisotropic strain due to current heating can generate this type of N{\'e}el vector reorientation \cite{Mee21}, as the N{\'e}el vector prefers alignment along an elongated $\langle 110 \rangle$ direction \cite{Gri22}. In principle, thermomagnetoelastic coupling and NSOT can cooperate for stripe and cross geometries aligned along $\langle 110 \rangle$ directions.

To study the action of the NSOT independently from this potentially additional contribution, we discuss the reversible current polarity dependence of the N{\'e}el vector reorientation, which cannot appear due to any thermally driven effect which acts identically for both current directions. For current pulses parallel to a hard $\langle 100 \rangle$-direction, the strongest NSOT is expected to act on the AFM domain walls, as only there the N{\'e}el vector is aligned parallel to the current. In this case the NSOT is expected to shift the domain wall reversibly with inverted current direction as observed previously for CuMnAs(001) thin films \cite{Wad18}. As discussed above and shown in Fig.\,3, we demonstrated such a current polarity dependent reorientation of the N{\'e}el vector as well. In Mn$_2$Au, a relatively large number of domains shows reversible modifications, but also in our compound the domain walls are shifted by distances of only $\approx \SI{1}{\micro m}$. This indicates that mechanisms such as domain wall pinning are competing with the NSOT acting on the domain walls. Nevertheless, the reversible current polarity dependent N{\'e}el vector reorientation represents direct evidence for the action of a current induced NSOT. 

In contrast to domain wall pinning, the potential barrier for N{\'e}el vector reorientation which is typically considered in theoretical work \cite{Zel14,Sel22} is the in-plane magnetocrystalline anisotropy ($\approx \SI{1.8}{\micro eV}$ per formula unit for Mn$_2$Au(001) \cite{Bom21}). In this framework a key role of thermal activation was previously reported based on investigations of granular Mn$_2$Au thin films \cite{Mei18}. However, as shown e.\,g.\,by the transmission electron microscopy image in the Supplementary Information, our Mn$_2$Au(001) thin films are highly epitaxial without morphological feature on the typical length scale of the AFM domains, so that independent switching of morphological grains can be excluded. This strongly suggests that the reorientation occurs via domain wall motion.

Independent of the physical origin of the potential barriers to be overcome, it is important for spintronics applications that the current induced N{\'e}el vector reorientation does not depend on slow thermal activation processes. Here, the required current density for full N{\'e}el vector reorientation using a single \SI{10}{\micro s} current pulse was only $\approx \SI{10}{\percent}$ larger than required for a pulse train of 100 current pulses with \SI{1}{\ms} pulse length each. This indicates that the current pulse induced force is at least of the same order of magnitude as opposing forces e.\,g.~due to domain wall pinning. 

We have shown that metallic antiferromagnets, specifically epitaxial Mn$_2$Au(001) thin films, are finally able to fulfill one of the major early promises of antiferromagnetic spintronics: it is possible to write long term stable information by single current pulses into macroscopic areas of device structures. Regarding the switching mechanism, while thermomagnetoelastic contributions might be present, we identify a N{\'e}el spin-orbit torque acting on the domain walls leading to reversible motion. Furthermore, we have shown that thermal activation processes are not essential for current induced N{\'e}el vector reorientation, thereby enabling fast and energy efficient switching. Thus, epitaxial Mn$_2$Au(001) thin films fulfill all requirements for memory applications regarding the writing and storage of data using a single antiferromagnetic layer.
 
\section{Methods}

All experimental data shown in this manuscript was obtained investigating Mn$_2$Au(001)(\SI{45}{\nm}) thin films grown epitaxially on Ta(001)(\SI{13}{\nm})/Mo(001)(\SI{20}{\nm}) double buffer layers on MgO(100) substrates. All layers were deposited by magnetron sputtering by the process described in detail in ref.\,\cite{Bom21}. The samples were capped with \SI{2}{\nm} of polycrystalline SiN$_x$ to protect them from oxidation. Optical lithography and ion beam etching were used to pattern the films into the cross and stripe structures shown in this manuscript.

Antiferromagnetic domain imaging was performed by combining photoemission electron microscopy with x-ray magnetic linear dichroism (XMLD-PEEM) at the Mn L$_{2,3}$ absorption edge at the PEEM endstations at beamline MAXPEEM at MAX IV, and beamline I06 at Diamond Light Source. The XMLD effect at the Mn L$_{2,3}$ absorption edge in Mn$_2$Au was established in previous work. For x-ray polarisation along a Mn$_2$Au $\langle 110 \rangle$ direction, the Mn L$_{2,3}$ XMLD spectrum shows a minimum and a maximum located at the absorption edge $E_{max}$ and at \SI{0.8}{\eV} below the edge. At MAXPEEM, the x-ray beam has normal incidence at the sample surface. The XMLD-PEEM images obtained at this beamline are the asymmetry images of images taken with two orthogonal x-ray polarisations along the $\langle 110 \rangle$ directions and energy of h$\nu$ = $E_{max}$-\SI{0.8}{\eV}, as this provides maximum contrast and minimum sensitivity to morphological features. 
At I06, the x-ray beam is incident at a grazing angle of \SI{16}{\degree}. XMLD-PEEM images measured there are the asymmetry of images with photon energies of h$\nu$ = $E_{max}$ and h$\nu$ = $E_{max}$ – \SI{0.8}{\eV} for fixed in-plane polarisation along a $\langle 110 \rangle$ direction.

In-situ electrical manipulation was performed using the pulse functions of Keithley2601B-PULSE (at MAX IV) and Keithley 2461 (at Diamond) sourcemeters, integrated into the X-PEEM setup.

The ex-situ resistance measurements (Fig.\,4) were performed using a Keithley 6220 precision current source with a probe current of 50~$\mu$A and a Keithley 2182A Nanovoltmeter in Delta mode averaging over 200 measurements to obtain one data point. For automatising the pulse (Keithley 2430 Pulse Source Meter) - probe sequence an Agilent 34970A Switch Unit was used.

For the quantification of the current pulse induced sample heating, we used a lithographic pattern, which additional to the cross shown in Fig.\,1 contains four thin leads enabling longitudinal resistance measurements of the central cross area with a 4-probe technique (see Fig.\,1 of ref.\,\cite{Bod18}). The maximum resistance obtained at the end of the current pulses with different lengths and amplitudes is shown in the Supplementary Information. By comparing this high pulse current sample resistance with the above room temperature linear extrapolated temperature dependent sample resistance $R(T)$ obtained from a low current measurement in a cryostat, we can deduce the current pulse induced temperature raise. These measurements were performed outside the PEEM at ambient conditions using the same Keithley 2601B-PULSE Source Meter as for pulsing in the PEEM. The time dependent sample resistance during the current pulses was measured with an oscilloscope and using the SENSE inputs of the Source Meter obtaining consistent results.   

\section{Supplementary Information}

{\bf Current pulse induced sample heating}

All current pulsing experiments are associated with ohmic heating of the sample. As Mn$_2$Au is a metal with an almost linear resistance versus temperature relation, the current pulse induced heating can be measured via a resistance measurement during the pulse as described in the Methods section. Fig.\,S1 shows the relative resistance increase of a Mn$_2$Au(001) sample patterned in a cross structure with a central section of $\SI{10}{\micro m}\times\SI{10}{\micro m}$ as in Fig.\,1 of the main text, but with additional leads to enable 4-probe resistance measurements. Different pulse lengths and amplitudes result in different heating. The dashed vertical lines in the main figure indicate the required current densities for complete switching of the N{\'e}el vector. From comparison with the extrapolated temperature dependent resistance measurement shown in the inset, the indicated temperatures associated with the switching using different pulse lengths are shown in the main panel.  

\begin{figure}
\includegraphics[width=0.9\columnwidth]{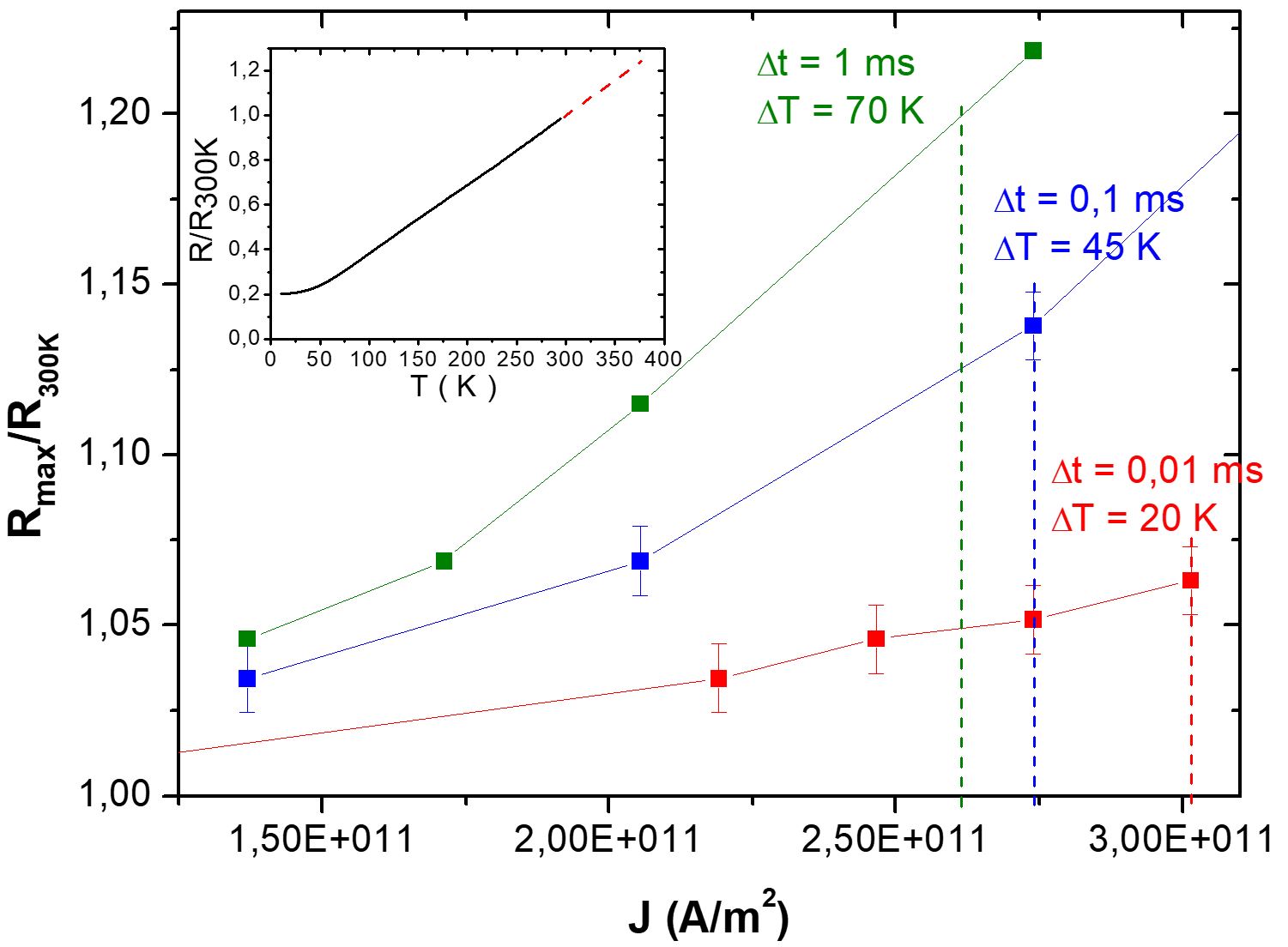}
\caption{\label{FigS1} 
{\bf S1: Current pulse induced sample heating.} The measurements in the main panel show the maximum sample resistance (normalized to the 300~K resistance) during application of current pulses with different length and amplitude. The dashed vertical lines show the required currents for full N{\'e}el vector reorientation corresponding to Fig.\,1. The inset shows the temperature dependence of the normalized resistance obtained from a standard dc measurement. Its extrapolation (red dashed line) is used to convert the resistance increase during current pulsing into a temperature increases $\Delta$T shown in the graph.} 
\end{figure}

{\bf Probing N{\'e}el vector reorientation by transverse resistance measurements}

Electrical detection of current pulse induced N{\'e}el vector alignment is also possible via measurements of the transverse resistance or planar Hall-Effect. Considering the rather small electrical signal, this is more easy to measure than the longitudinal AMR, as there is no offset voltage. In Fig.\,S2, the measurement of the transverse resistance changes is shown in analogy to Fig.\,4 of the main text.

\begin{figure}
\includegraphics[width=0.9\columnwidth]{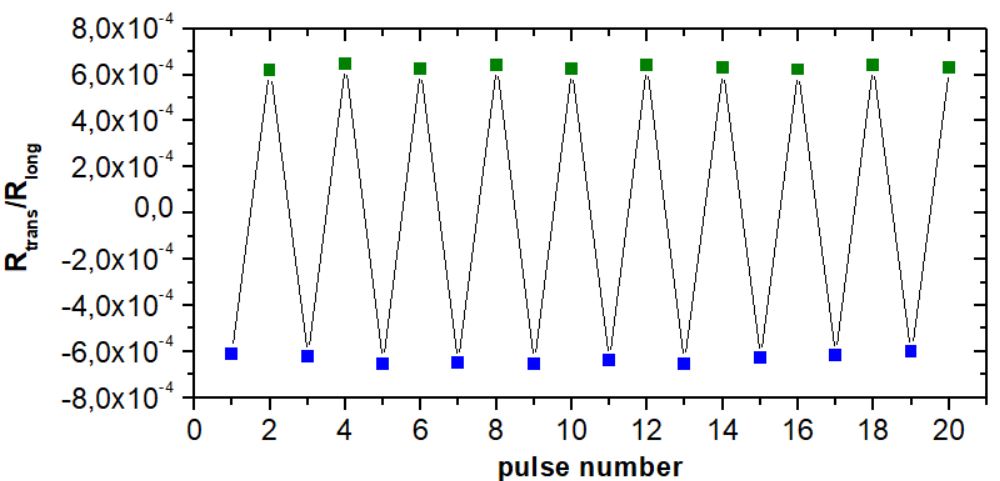}
\caption{\label{FigS2} 
{\bf S2: Switching induced planar Hall effect.} The measurements shows the alternating transverse resistance
of a Mn$_2$Au thin film patterned in the geometry shown in Fig.\,4, panel {\bf b}. The transverse resistance was measured after each 1~ms current pulse with $J_{\rm pulse}=3 \times 10^{11}$~A$/$m$^2$ applied alternately in perpendicular directions as indicated by the blue and green arrows in Fig.\,4, panel {\bf b}. The color of the data points corresponds to the arrows indicating the current direction.} 
\end{figure}

{\bf Stability of the N{\'e}el vector alignment}

All images of current induced N{\'e}el vector reorientation shown in the main text were obtained at MAXPEEM (MAX~IV). Four months later at  Diamond Light Source, we were able to image the sample shown in Fig.\,2 of the main text again. As shown in Fig.\,S3, we observed exactly the same domain pattern again, i.\,e.\,the current induced N{\'e}el vector oriented domain configuration obtained four month ago was completely preserved.

\begin{figure}
\includegraphics[width=0.8\columnwidth]{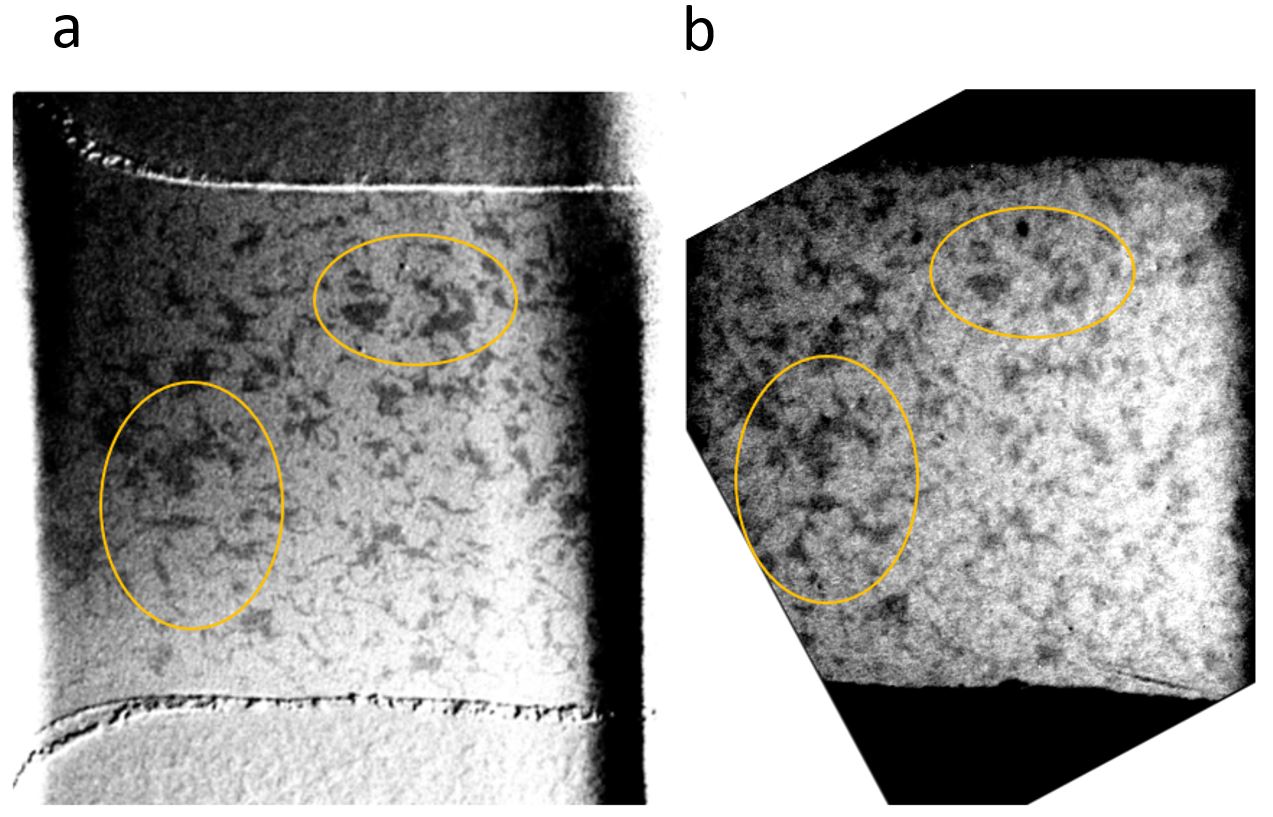}
\caption{\label{FigS3} 
{\bf S3: Long term stability of switched state.} The XMLD-PEEM image in panel {\bf a} is the same as shown in Fig.\,2, right panel, obtained after current induces N{\'e}el vector reorientation. The image in panel {\bf b} shows the same area 4 months later. The yellow circles serve to guide the eye showing exactly the same AFM domain pattern (please note that the image in panel {\bf b} has reduced magnetic contrast due to a different PEEM geometry).} 
\end{figure}

Additionally, we demonstrate the absence of relaxation on short time scales, i.\,e.\,seconds, by measurements of the transverse resistance after a sequence of current pulse induced reversible switching. The data shown in Fig.\,S4 was obtained in the same way as for Fig.\,S2 (here with $J_{\rm pulse}=2.9 \times 10^{11}$~A$/$m$^2$), but we kept measuring the resistance after the last switching current pulse for $1.5$~h. The raw data shown was obtained at a sampling rate of $\simeq 2$ data points per second. Also on this time scale, no relaxation of the resistance after the last switching pulse was observed.

\begin{figure}
\includegraphics[width=0.8\columnwidth]{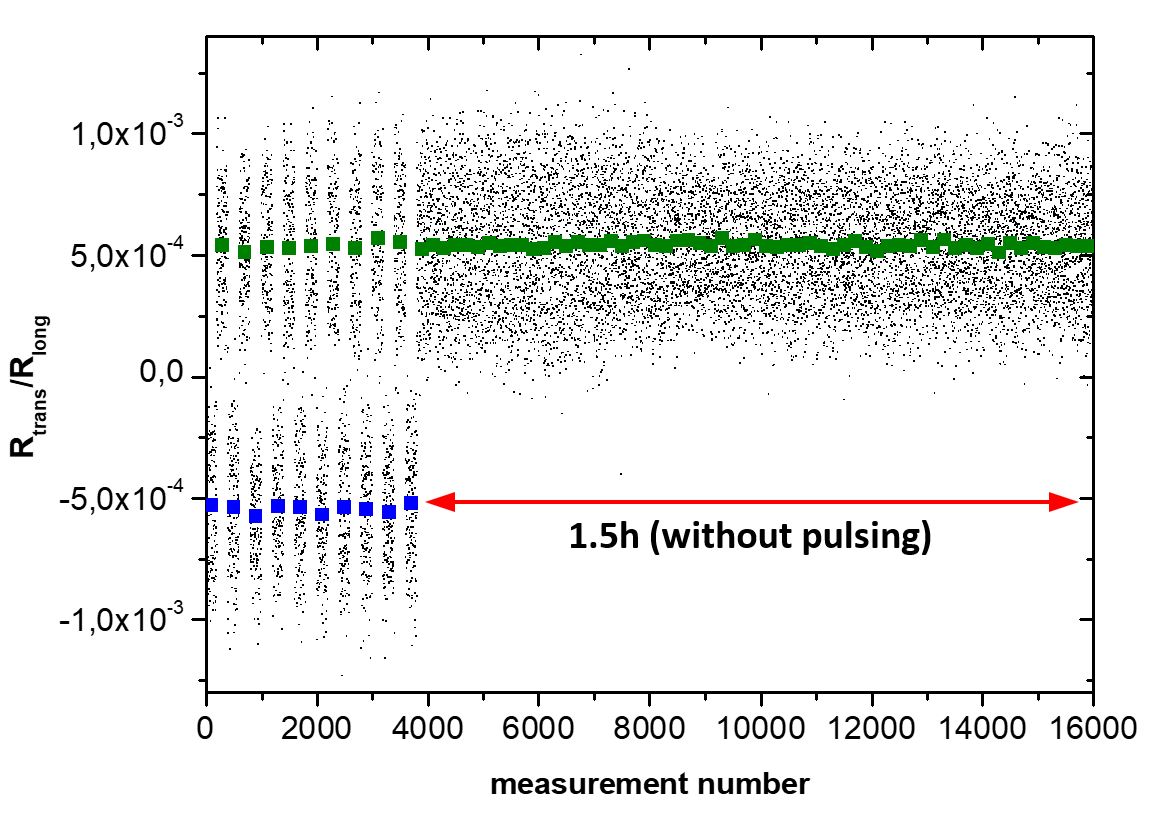}
\caption{\label{FigS4} 
{\bf S4: Second time-scale stability of switched state.} Raw data (black data points) and averaged data points (over 200 raw data points) of the transverse resistance of a Mn$_2$Au thin film. Here, the transverse resistance was measured after each 1~ms current pulse with $J_{\rm pulse}=2.9 \times 10^{11}$~A$/$m$^2$ applied alternately in perpendicular directions followed by pure resistance measurements without pulsing. No relaxation was observed.} 
\end{figure} 

{\bf Microstructure of Mn$_2$Au(001) thin films}

All epitaxial Mn$_2$Au thin films used for this manuscript were prepared by sputtering as described in the Methods sections. Beyond their characterization shown in ref.\,[20], we here want to emphasize their long range crystallographic order by showing in Fig.\,S5 scanning transmission electron microscopy images. Panel {\bf a} with the highest magnifications shows the Mn$_2$Au lattice with atomic resolution (on top of this sample is a Permalloy layer, which was not present in the other samples discussed here.). Panel {\bf b} with a larger field of view shows a MnAu impurity phase at the interface with the Ta buffer layer, which, however, does not disturb the growth of the Mn$_2$Au layer on top of it. Panel {\bf c} shows the largest field of view demonstrating the long range coherence of the epitaxial growth.

\begin{figure}
\includegraphics[width=0.9\columnwidth]{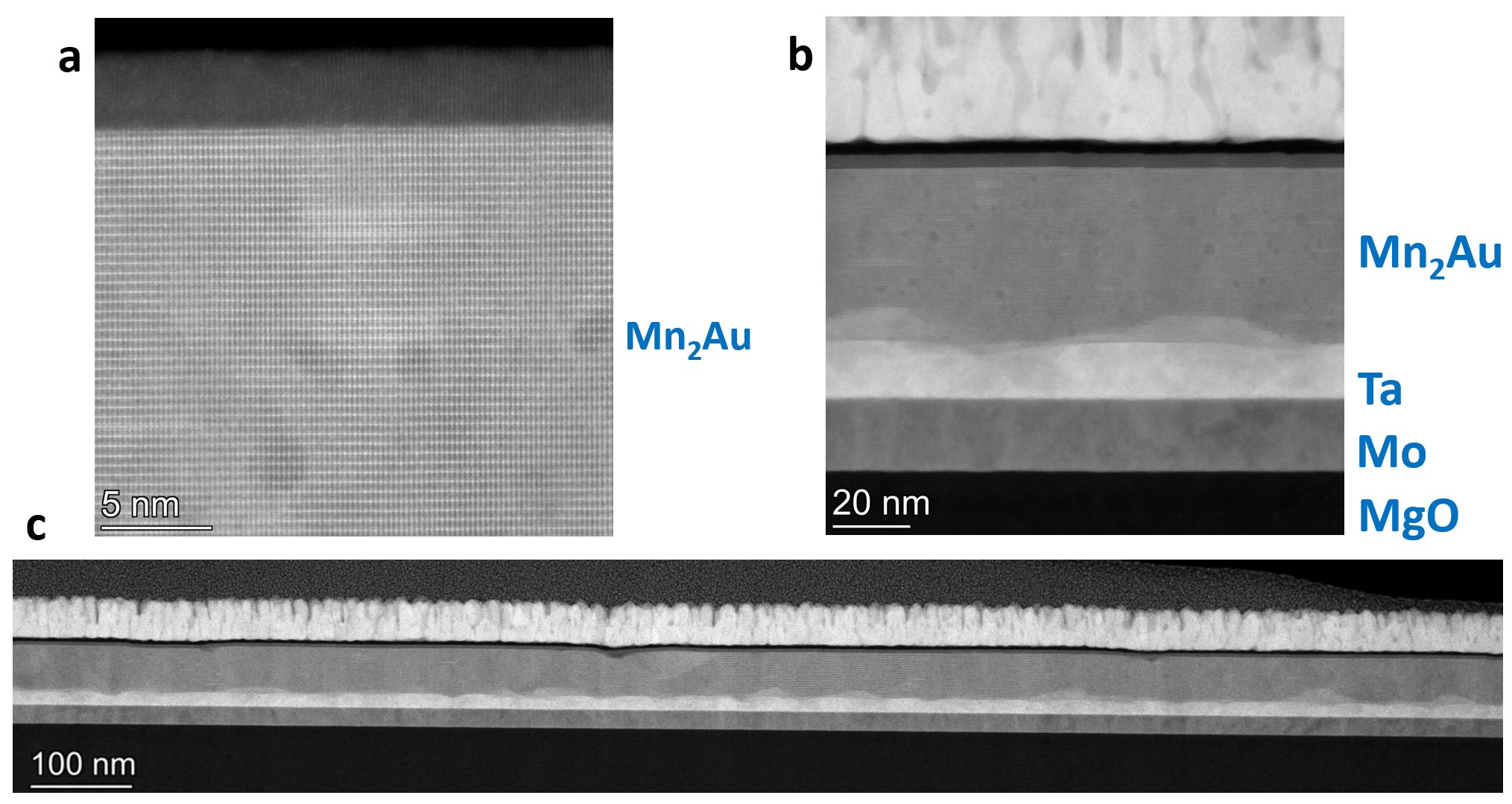}
\caption{\label{FigS4} 
{\bf S5: Microstructure of a Mn$_2$Au(001) epitaxial thin film.} Cross section HAADF STEM images with different magnifications showing the microstructure of an epitaxial Mn$_2$Au(001) thin film. A weak MnAu impurity phase is visible at the interface between the Mn$_2$Au and the Ta buffer layer. The samples show long range crystallographic order.} 
\end{figure}

For obtaining the STEM images, an electron transparent cross-section lamella was prepared using a \SI{30}{\kV} focused Ga+ ion beam and scanning electron microscope (FIB-SEM) FEI Helios platform. The ion beam energy was decreased to 5 kV for the final thinning steps. Scanning transmission electron microscopy (STEM) was carried out using an FEI Titan TEM equipped with a Schottky ﬁeld emission gun operated at \SI{200}{\kV}, a CEOS probe aberration corrector and a high angle annular dark-ﬁeld detector (HAADF).

{\bf Data availability}

The raw data of the transport measurements shown in this study has been deposited in the Zenodo database under accession code DOI 10.5281/zenodo.7467238.

{\bf Acknowledgements}

We acknowledge funding by the Deutsche Forschungsgemeinschaft (DFG, German Research Foundation) - TRR 173 - 268565370 (project A05, with contribution from A01), by the Horizon 2020 Framework Program of the European Commission under FET-Open Grant No. 863155 (s-Nebula), by EU HORIZON-CL4-2021-DIGITAL-EMERGING-01-14 programme under grant agreement No. 101070287 , and by the TopDyn Center. We acknowledge MAX IV Laboratory for time on beamline MAXPEEM under Proposal 20210863, Diamond Light Source for time on beamline I06 under proposal MM30141-1, and Swiss Light Source for time on beamline SIM. Research conducted at MAX IV, a Swedish national user facility, is supported by the Swedish Research council under contract 2018-07152, the Swedish Governmental Agency for Innovation Systems under contract 2018-04969, and Formas under contract 2019-02496. The STEM investigations were funded by the European Union's Horizon 2020 Research and Innovation Programme under grant agreement 856538 (project “3D MAGIC”).

{\bf Author contributions}

S.R., Y.L., and M.J. wrote the paper, prepared the samples and performed the XMLD-PEEM investigations. Y.R.N and E.G. supported the XMLD-PEEM investigations at MAX IV, B.S. and L.S.I.V. at Diamond, and A.K. at SLS. S.R. and J.B. measured the N{\'e}el vector induced resistance changes, Y.R.N. measured the pulse heating  induced resistance changes. T.D., A.K. and R.E.D-B contributed the STEM investigations. M.K. contributed to the discussion of the results and provided input, M.J. coordinated the project.

{\bf Competing interests:}

 The authors declare no competing financial interests.

\end{document}